\documentclass[10pt,twocolumn, aip,cha,letterpaper,groupedaddress,showpacs,amsmath,amssymb,floatfix]{revtex4-1}

\usepackage{color}
\usepackage{graphicx}
\usepackage{graphics}
\usepackage{natbib}
\usepackage{setspace}

\begin{document}
\title{A pseudo-matched filter for chaos}
\author{Seth D. Cohen and Daniel J. Gauthier}
\affiliation{Department of Physics, Duke University, Durham, North Carolina 27708, USA}
\date{\today}

\begin{abstract}
A matched filter maximizes the signal-to-noise ratio of a signal. In the recent work of Corron \textit{et al.} [Chaos {\bf 20}, 023123 (2010)], a matched filter is derived for the chaotic waveforms produced by a piecewise-linear system. Motivated by these results, we describe a pseudo-matched filter, which removes noise from the same chaotic signal. It consists of a notch filter followed by a first-order, low-pass filter. We compare quantitatively the matched filter's performance to that of our pseudo-matched filter using correlation functions in a simulated radar application. On average, the pseudo-matched filter performs with a correlation signal-to-noise ratio that is 2.0 dB below that of the matched filter. Our pseudo-matched filter, though somewhat inferior in comparison to the matched filter, is easily realizable at high speed ($>$ 1 GHz) for potential radar applications.
\end{abstract}

\maketitle
\begin{quotation}
Radar measures the propagation times of transmitted electromagnetic waves between a target and a receiver. A stored copy of the transmitted waveform is compared to the received signal to compute relative distances and locate objects. The resolution of a radar is proportional to the bandwidth of the transmitted signal and, in the case of extremely large bandwidth, the radar must be capable of storing and processing large amounts of data. One way to maintain high bandwidth and circumvent this sampling limitation is to use chaos. Chaotic systems are deterministic, non-repeating, and can be engineered to have an underlying symbolic dynamics. Rather than sampling and storing the entire chaotic waveform in a radar, one can track the symbols associated with the transmitted signal. In this technique, a receiver must recover the symbolic dynamics from the received signal, which has picked up noise from the scattering environment. Recently, Corron \textit{et al.} developed a piecewise-linear chaotic system that uses a matched filter to recover the system's symbolic dynamics. Using these results, we empirically derive a simpler \textit{pseudo-matched} filter that uses high-speed commercially available filters to recover the system's symbolic dynamics. 
\end{quotation}

\section{Introduction}
\begin{figure}[t!]
\begin{center}
\includegraphics[width=80mm]{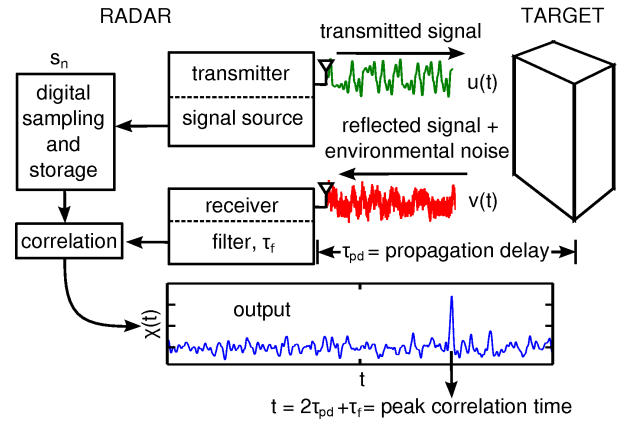}
\end{center}
\caption{\label{fig:Radar_Concept}Example chaos radar. A signal $u(t)$ is transmitted to measure target distances and stored digitally as $s_n$. The received signal $v(t)$ is filtered and correlated with $s_n$. The correlation output $\chi(t)$ peaks at time $t = 2\tau_{\text{pd}} + \tau_{\text{f}}$, where $\tau_{\text{pd}}$ is the propagation delay from the radar to the target and $\tau_{\text{f}}$ is the delay through the filter.}
\end{figure}

A conventional radar system measures the distances of targets in the field of view using a signal source, a transmitter, and a receiver. In Fig.~\ref{fig:Radar_Concept}, a radar transmitter broadcasts a signal $u(t)$ from the source toward an intended target, and the receiver detects a version of the transmitted signal that is reflected off of the target. Prior to transmission, a copy of the radar's signal is digitally sampled and stored as $s_n$. The received signal $v(t)$, which picks up environmental noise, is filtered and correlated with $s_n$. Typical radar signals are non-repeating in order to avoid multiple points of correlation and, therefore, the correlation will peak only when the transmitted and received signals are aligned. Using the time of the output peak in the correlation, the measured range from the transmitter to the target is determined. 

The performance of a radar is determined by its ability to identify the correlation time between the transmitted and received signal. In the correlation function, the width of the peak scales inversely with the bandwidth of the transmitted signal and sets the spatial resolution of the radar. In addition, the height of the correlation peak above the noise floor, also known as signal-to-noise ratio ($SNR$), is determined by the length of the transmitted and sampled waveforms as well as the noise from the environment. Thus, the digital storage capacity of the radar sets the maximum $SNR$ in the correlation measurement. State-of-the-art radar systems use high-frequency, broadband signals, where the digital sampling and storage of the signals can be costly. These radars must balance the bandwidth and cost of the system’s design while maintaining its performance.

A simple example of an inexpensive, broadband signal source is amplified electrical noise. In the past, electrical noise  has been used by radar systems to perform ranging measurements.\cite{Liu2003,T2010} The high bandwidth of these noise generators yields high-resolution ranging information, but requires fast sampling and large data storage capacities. Some recent techniques have been proposed to minimize the necessary sampling capacity of noise radars using analog delay lines for signal storage.\cite{Lukin2010} But, these methods limit the ranging capabilities of the radar. Further research has shown that more efficient radars benefit from a structured, rather than a stochastic, signal.

For instance, various deterministic signal sources have been studied in efforts to minimize the necessary data storage rate and capacity of a radar. As one example, pseudo-random binary sequences (PRBS's) are often used as a radar signal sources. To be implemented as a radar signal source, a PRBS is up-converted to a suitable frequency band before transmitting and then down-converted at the receiver before correlation.\cite{Uraz2007,Mirsha2012} The main advantage of a PRBS is the ability to use one-bit digital samplings of the binary sequences, thereby requiring low amounts of data-storage capacity. This allows for longer sequences of the transmitted waveform to be stored, thus enhancing the radar's $SNR$ without increasing the cost of the system. The main disadvantage of a PRBS is that it requires computational power to generate and its sequence eventually repeats, which ultimately limits its performance. Many other radar concepts like this one exist, each with advantages and disadvantages, and today the radar community continues to develop broadband signal sources.\cite{Dmitriev2007,Khan2010}

One novel approach to a radar is to use chaotic waveform generators, which are believed to have several properties that make them ideally suited as signal sources for radar applications.\cite{Sobhy2000} One defining feature of a chaotic system is that it can generate a signal that does not repeat in time. Chaotic signals are also often inherently broadband. High-speed chaos has been observed in optical and electronic systems with frequency bandwidths that stretch across several gigahertz.\cite{Kouomou2005,Illing2006, Zhang2009} Such broadband chaos has been studied for its applications in high-resolution ranging and in imaging.\cite{Myneni2001,Lin2004, Venk2005, Venk2009} These applications use the non-repeating aspects of the high-speed chaos.

However, conventional chaos radars do not take full advantage of the chaotic signal source. In addition to producing broadband, non-repeating signals, chaotic systems are deterministic and extremely sensitive to small perturbations. By not using these properties, chaos radars add no benefit over noise radars, requiring high-sampling to perform correlations. Thus, to take full advantage of chaotic systems in a radar application, a chaos radar needs to benefit from the determinism or sensitivity of the chaos in addition to its noise-like properties.

Recently, Corron \textit{et al.} proposed a novel chaos radar concept that uses the dynamics produced by a piecewise-linear harmonic oscillator.\cite{Saito1981, Corron2010, Blakely2010} It produces simultaneously a chaotic signal and a binary switching state that completely characterizes its chaotic dynamics. In the proposed radar, the chaotic signal serves as the signal source (see Fig.~\ref{fig:Radar_Concept}) and is transmitted, while a copy of a switching state is stored using a one-bit digital sampling. For a radar receiver, Corron \textit{et al.} derived the analytical form of a filter that is matched to the chaos produced by the system. A matched filter is a linear operation that optimizes the $SNR$ of a signal in the presence of additive white Gaussian noise ($AWGN$). Their matched filter recovers the switching states from the received signal, which are then passed to the correlation operation of Fig.~\ref{fig:Radar_Concept}. This technique uses deterministic aspects of the chaos, making it an improvement over a noise radar. With reduced data storage and an optimal $SNR$, their architecture could effectively reduce the cost of a radar system. 

Corron \textit{et al.} implement their piecewise-linear design using an LRC (inductance-resistance-capacitance) oscillator that operates in the kHz frequency range.\cite{Corron2010} It is difficult to a realize high-frequency version ($>$ 1 GHz) of this system because of parasitic capacitances, inductances, and the inherent time delays in the propagation of signals in LRC circuits. As it stands, there is no high-frequency realization of the piecewise-linear system from Ref. [\onlinecite{Corron2010}] or the associated matched filter. In order to have resolutions that are comparable to state-of-the art radar systems, the waveforms and switching states produced by this chaotic system must be scaled to higher-frequencies and to broader bandwidths. Thus, techniques for increasing the speed of this system are of interest. 

As a first step toward a high-frequency implementation of Corron \textit{et al.'}s system, we compare the performance of the matched filter presented in Ref. [\onlinecite{Corron2010}] to a set of standard filters (first order low-pass filter and notch filter). These standard filters operate at high-frequencies, are inexpensive, well characterized, and readily available. Cascading these standard filters allows us to realize a pseudo-matched filter for the chaos produced by the piecewise-linear harmonic oscillator. We define a pseudo-matched filter as a sub-optimal linear operation (when compared to the matched filter) that removes $AWGN$ from the corresponding (matched) waveform and performs comparably to the matched filter for the system. We are interested in a chaos-based radar system such as that described by Ref. [\onlinecite{Blakely2010}] that could benefit from integrating readily available filters for high-speed applications.

\section{Matched Filter Review}

To motivate our analysis, we briefly review the characteristics of the chaotic system and matched filter presented in Ref. [\onlinecite{Corron2010}] within the context of a radar application. Consider a harmonic oscillator with negative damping $-\beta$ and with a piece-wise constant driving term $s(t)$ whose behavior is governed by the differential equation
\begin{equation}
\ddot {u}(t) - 2 \beta \dot{u}(t)+ (\omega_{\text{o}}^2+\beta^2)(u(t)-s(t)) = 0,
\label{eq:harmonic_oscillator_1}
\end{equation}
together with a guard condition on the output variable $u(t)$ that switches the sign of $s(t)$ 
\begin{equation}
s(t) = \left\{
     \begin{array}{lr}
       \;\;\,1, & \text{if } u(t^*) \geq 0 \\
       -1, & \text{if } u(t^*) < 0
     \end{array}
   \right..
\label{eq:harmonic_oscillator_2}
\end{equation}

\begin{figure}[t!]
\begin{center}
\includegraphics[width=70mm]{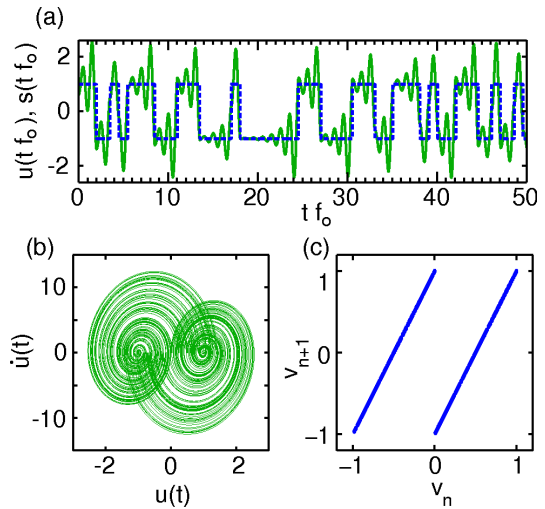}
\end{center}
\caption{\label{fig:US_TimeSeries}Chaos from a piecewise-linear harmonic oscillator with negative damping. (a) Time series of the analog variable $u(t)$ (green) and the nonlinear switching state $s(t)$ (blue dashed line). (b) Chaotic attractor in phase space. (c) Chaotic shift map created by sampling $u(t)$ using the times $t^*$ from Eq. (2), where $v_n = u(t^*)$ if $|u(t^*) - 1| < 0$.}
\end{figure}

Figure~\ref{fig:US_TimeSeries} shows a time series of the variables $u(t)$ and $s(t)$ with the parameters $\omega_{\text{o}} = 2\pi$ and $\beta = ln(2)$. We integrate Eqs. (\ref{eq:harmonic_oscillator_1})-(\ref{eq:harmonic_oscillator_2}) using MATLAB's ODE45, where the switching condition is monitored using the integrator's event-detection algorithm. The attractor for this system is low dimensional and is plotted in Fig.~\ref{fig:US_TimeSeries}b. The dynamics can also be viewed as a chaotic shift map,\cite{Corron2010} as seen in Fig.~\ref{fig:US_TimeSeries}c. For this system, $u(t)$ oscillates with a growing amplitude and fixed oscillation frequency $f_{\text{o}} = \omega_{\text{o}} /2\pi$ about a piece-wise constant line $s(t)$. The switching times of $s(t)$ depend on the local maxima and minima of $u(t)$, and the times between the local maxima and minima of $u(t)$ are fixed by the fundamental frequency $f_{\text{o}}$. Thus, the maximum rate of switching in $s(t)$ is limited to $1/f_{\text{o}}$. Using a one-bit digital sampling of $s(t)$ at a sampling frequency that is greater than or equal to $f_{\text{o}}$, we are able to store a record of all switching state values $s_n$. Similar to the case for a PSRB, information about the transmitted waveform can be stored with minimal sampling and memory, enhancing the potential $SNR$ of a radar correlation measurement. 

\begin{figure}[t!]
\begin{center}
\includegraphics[width=80mm]{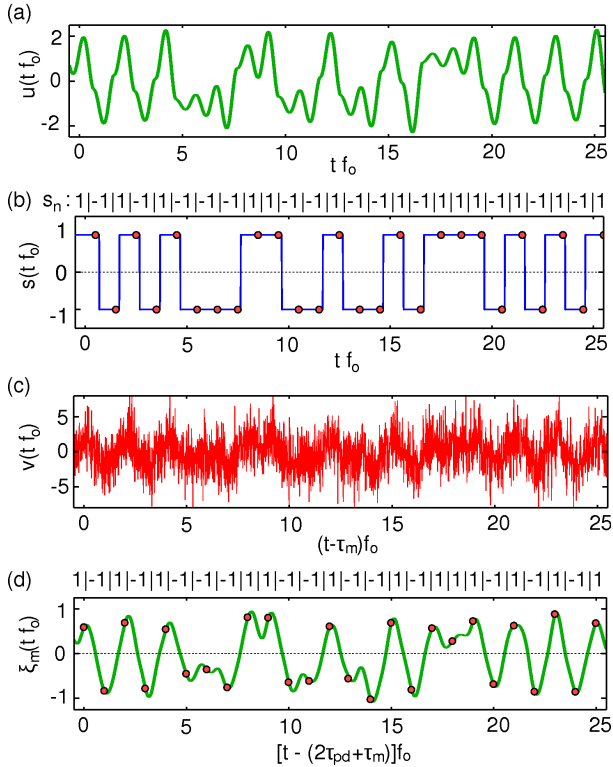}
\end{center}
\caption{\label{fig:Matched_Filter_Time_Series}Temporal evolution of (a) $u(t)$, (b) $s(t)$, (c) $v(t)$ with $SNR = -5.9$ dB, and (d) $\xi_{\text{m}}(t)$. The horizontal axes in (b) and (c) are shifted by $\tau_{\text{m}}$ and $(2\tau_{\text{pd}} + \tau_{\text{m}})$, respectively, where $\tau_{\text{pd}}$ is the propagation distance to an intended target and $\tau_{\text{m}}$ is the time delay through the matched filter. Above the signals $\xi_{\text{m}}(t)$ and $s(t)$ (blue), a single-bit discrete sampling of the waveforms (red dots) is shown.}
\end{figure}

In addition to its data storage capabilities, chaos from this system can be further exploited using a matched filter. Corron \textit{et al.} demonstrate that the switching information $s(t)$ can be recovered using a matched filter at the  radar receiver.\cite{Blakely2010} The match filter is described by
\begin{equation}
\dot{y}(t) = v(t-2\pi/\omega_{\text{o}})-v(t),
\label{eq:matched_filter_time_domain_1}
\end{equation}
\begin{equation}
\ddot{\xi_m}(t)+2 \beta \dot{\xi_m}(t)+ (\omega_{\text{o}}^2+\beta^2)\xi_m(t) = (\omega_{\text{o}}^2+\beta^2)y(t),
\label{eq:matched_filter_time_domain_2}
\end{equation}
where $v(t)$ is the input signal and $\xi_{\text{m}}(t)$ is the analog output of the matched filter. In Fig.~\ref{fig:Matched_Filter_Time_Series}, we examine the output of the matched filter when it is driven by $v(t) = u(t)$ + $AWGN$. The original signals $u(t)$ and $s(t)$ are plotted in Figs.~\ref{fig:Matched_Filter_Time_Series}a-b. The switching state $s(t)$ is plotted with a one-bit digital sampling $s_n$ at a sampling frequency $f_{\text{o}}$. In Fig.~\ref{fig:Matched_Filter_Time_Series}c, the time series $v(t)$ has a $SNR$ of $-5.9$ dB, where $SNR = 10\text{log}_{10}(SNR_{\text{input}})$, $SNR_{\text{input}} = \sigma_{u}^2/\sigma_{AWGN}^2$, and where $\sigma_{u}^2$ and $\sigma_{AWGN}^2$ are the input signal $u(t)$ and additive noise powers, respectively (see the Appendix for details on the additive noise). The matched filter is optimized to remove the noise from $v(t)$ and recover the binary switching information $s_n$ for the correlation operation.

We compare $s_n$ to the solution for Eqs. (\ref{eq:matched_filter_time_domain_1}) - (\ref{eq:matched_filter_time_domain_2}), $\xi_{\text{m}}(t)$, shown in Fig.~\ref{fig:Matched_Filter_Time_Series}d. The solution $\xi_{\text{m}}(t)$ is the output of the matched filter and is a nearly noise-free waveform that transitions approximately between two states, one positive and negative (defined by a dotted black line at $\xi_{\text{m}}(t) = 0$). In between these transitions, $\xi_{\text{m}}(t)$ oscillates with a small, irregular amplitude at a frequency approximately equal to $f_{\text{o}}$. Upon further examination, the matched filter output signal $\xi_{\text{m}}(t)$ follows the same transitions from the original switching state $s(t)$. Using a correlation between $s(t)$ and $\xi_{\text{m}}(t)$, we determine the time delay through the matched filter $\tau_{\text{m}}$ to be approximately $1/(2f_{\text{o}})$. We compensate for the delay, and sample $\xi_{\text{m}}(t)$ at $f_{\text{o}}$, assigning binary values using the relation: $-1$ if $\xi_{\text{m}}(t_n) \leq 0$ and $+1$ if $\xi_{\text{m}}(t_n) > 0$, where $t_n$ is the $n^{\text{th}}$ sampling time. These binary values, shown in Fig.~\ref{fig:Matched_Filter_Time_Series}d, match $s_n$, demonstrating that the matched filter has recovered the switching information from $v(t)$.

In a simulated radar application, Corron \textit{et al.} use a tapped delay line to perform a time-domain correlation operation between the digitally stored $s_n$ and the analog output of the matched filter $\xi_{\text{m}}(t)$. The tapped delay line is described by
\begin{equation}
\chi_{\text{m}}(t) = \sum\limits_{n=1}^N s_n \xi_{\text{m}}(t-\tau_n),
\label{eq:correlation_matched}
\end{equation}
where $N$ is the length of the stored sequence and $\chi_{\text{m}}(t)$ is the output of the correlator (we have used the linearity of Eqs. (\ref{eq:harmonic_oscillator_1}) - (\ref{eq:matched_filter_time_domain_2}) to rearrange the operations in Ref. [\onlinecite{Blakely2010}]). The tapped delay line is a convenient technique for calculating in real time the correlation between a binary and an analog signal. 

\begin{figure}[t!]
\begin{center}
\includegraphics[width=85mm]{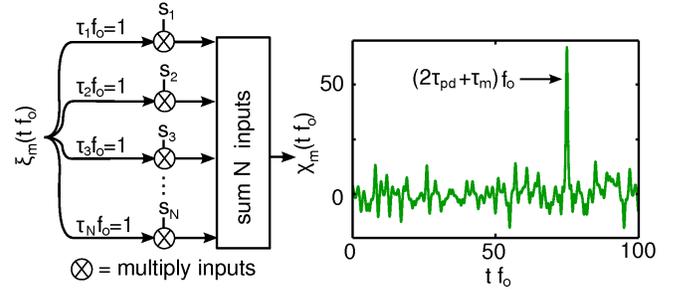}
\end{center}
\caption{\label{fig:Tapped_Delay_Line}Schematic representation of the tapped delay line in Eq. (\ref{eq:correlation_matched}). The switching state $s(t)$ is sampled and $s_n$ is stored for $n = 1$ to $n = N$, where, $N=100$ in this case. The output of the matched filter $\xi_{\text{m}}(t)$ drives the tapped delay line and the output sum $\chi_{\text{m}}(t)$ peaks at time $t = 2\tau_{\text{pd}}+\tau_{\text{m}}$.}
\end{figure}

To better understand the operations performed by a tapped delay line, we use a pictorial representation of Eq. (\ref{eq:correlation_matched}). In Fig.~\ref{fig:Tapped_Delay_Line}, the output of the matched filter $\xi_{\text{m}}(t)$ splits into $N$ copies, each of which are successively delayed by $\tau_n f_{\text{o}} = n$, where $n$ is an integer. The resulting copies are multiplied by the corresponding stored $s_n$ and summed continuously in time. The output $\chi_{\text{m}}(t)$ peaks $2\tau_{\text{pd}} + \tau_{\text{m}}$, where $\tau_{\text{pd}}$ and $\tau_{\text{m}}$ are the propagation delays of the signal to the target and through the matched filter, respectively. In the context of Fig.~\ref{fig:Radar_Concept}, the correlation operation of this chaos radar can be performed between a transmitted and received signal using the tapped delay line in Eq. (\ref{eq:correlation_matched}). Thus, using the chaotic waveform generator from Ref. [\onlinecite{Corron2010}], combined with the matched filter and tapped delay line, we arrive at a chaos radar.\cite{Blakely2010} 

This particular chaos radar benefits in two ways from the deterministic characteristics of the system's chaos. The first benefit is the link between the non-repeating waveform $u(t)$ and the switching state $s(t)$: A binary sampling of $s(t)$ can completely characterize the dynamics in $u(t)$. A second benefit is the ability to derive a matched filter that optimizes the output $SNR$ of the receiver. The matched filter for chaos, combined with a tapped delay line, provides an architecture for relatively quick and inexpensive correlations between a binary switching signal and a recovered analog signal, both generated from a single chaotic system. These benefits present a platform for a high-performance radar with improved $SNR$ over conventional chaos radars. 

\section{Matched Filter Analysis}

With this background and motivation, we now examine Corron \textit{et al.}'s matched filter for chaos in the frequency domain. Using the transfer functions of Eqs. (\ref{eq:harmonic_oscillator_1}), (\ref{eq:matched_filter_time_domain_1}), and (\ref{eq:matched_filter_time_domain_2}), we examine the spectral properties of the matched filter. For the purposes of our analysis, we focus on the magnitudes of transfer functions. Because the phase of the matched filter is approximately linear with frequency $f$ when $f < f_{\text{o}}$, it preserves the transition information in $s_n$. We will derive empirically a pseudo-matched filter using a combination of standard filters and that also preserves the transition information in $s_n$. Lastly, we compare the pseudo-matched filter performance to the true matched filter for chaos in a simulated radar application.

\begin{figure}[t!]
\begin{center}
\includegraphics[width=65mm]{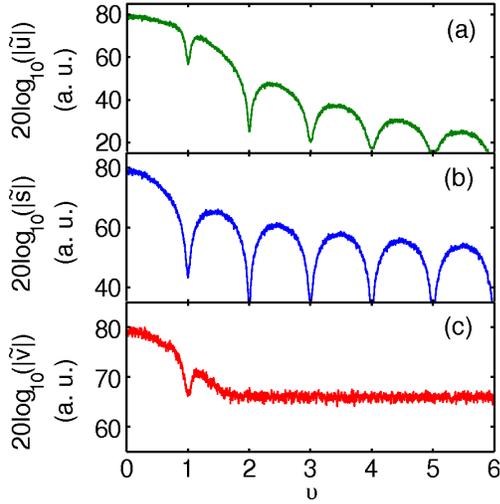}
\end{center}
\caption{\label{fig:Frequency_Spectra}Frequency spectra for the chaotic time series of (a) $u(t)$, (b) $s(t)$, and (c) $v(t)$. The frequency spectra are calculated using $\tilde{u}$, $\tilde{s}$, and $\tilde{v}$, the discrete Fourier transforms of $u(t)$, $s(t)$, and $v(t)$, respectively. In these plots, the frequency axes have a spectral resolution of $10^{-4}$, and the spectral amplitudes have been averaged over a window of $10^{-2}$.}
\end{figure}

First, we analyze the spectral properties of the chaotic dynamics from $u(t)$ and $s(t)$ as well as the driving signal $v(t)$. In Fig.~\ref{fig:Frequency_Spectra}, we plot the spectral amplitudes for $u(t)$, $s(t)$, and $v(t)$, where $v(t) = u(t)$ + $AWGN$ ($SNR = -5.9$ dB). One should take note that, in Fig.~\ref{fig:Frequency_Spectra}a, there is no maxima in the frequency spectrum at $f_{\text{o}}$, the fundamental frequency of oscillation, because the phase of $u(t)$ switches by $\pi$ each time $s(t)$ switches states. This demonstrates that, if the spectrum of $u(t)$ is scaled-up in high-frequency system ($> 1$ GHz), the bandwidth in $u(t)$ would stretch over several gigahertz. In addition to its radar properties, a high-frequency broadband spectrum in $u(t)$ would provide a useful carrier signal for low-profile or ultra-wideband technologies.\cite{Withington2003, Volkovskii2005} In Fig.~\ref{fig:Frequency_Spectra}c, the frequency spectrum of the $AWGN$ in $v(t)$ covers information about $u(t)$. The matched filter is engineered to process the spectrum of $v(t)$ to recover $s_n$.

\begin{figure}[t!]
\begin{center}
\includegraphics[width=65mm]{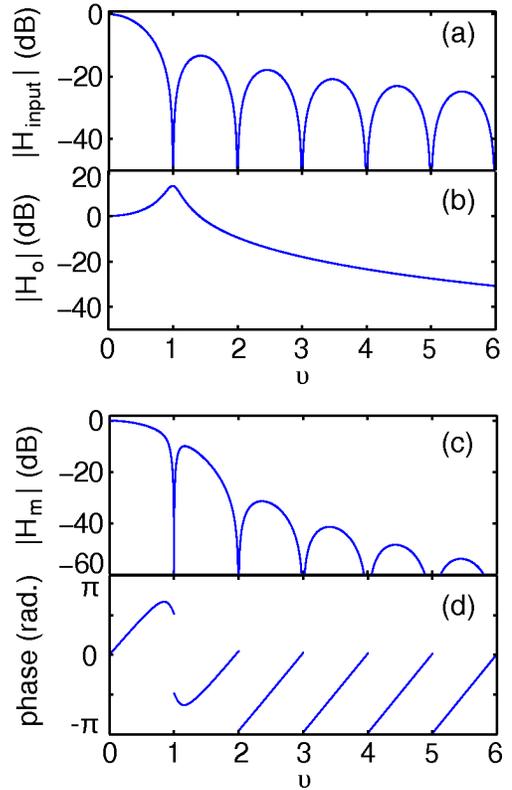}
\end{center}
\caption{\label{fig:Matched_Filter_Transfer_Functions}Magnitudes the transfer functions (a) $H_{\text{input}}$, (b)  $H_{\text{o}}$, and (c) $H_{\text{m}}$. In (d), the phase of $H_{\text{m}}$ is given as a function frequency.}
\end{figure}

Next, we examine the transfer function of the matched filter. We take the Fourier transform of Eqs. (\ref{eq:matched_filter_time_domain_1}) - (\ref{eq:matched_filter_time_domain_2}) and obtain the transfer functions $H_{\text{input}}$ and $H_{\text{o}}$. We combine these transfer functions to obtain the transfer function for the matched filter $H_{\text{m}}$. They read
\begin{equation}
H_{\text{input}}(\nu) = \frac{\tilde{y}}{\tilde{v}} = \frac{e^{2\pi i \nu}-1}{2 \pi i \nu},
\label{eq:H_input}
\end{equation}
\begin{equation}
H_{\text{o}}(\nu)= \frac{\tilde{\xi}_{\text{m}}}{\tilde{y}} = \frac{4 \pi^2 + \beta^2}{4\pi^2(1-\nu^2)+\beta(4\pi i \nu + \beta)},
\label{eq:H_dynamical}
\end{equation}
\begin{equation}
H_{\text{m}}(\nu) = \frac{\tilde{\xi}_{\text{m}}}{\tilde{v}} = H_{\text{input}} H_{\text{o}},
\label{eq:H_matched}
\end{equation}
where $\nu = f/f_{\text{o}}$ and $\tilde{\xi}_{\text{m}}$, $\tilde{y}$, and $\tilde{v}$ are the Fourier transforms of $\xi_{\text{m}}(t)$, $y(t)$, and $v(t)$, respectively. We plot the magnitudes of $H_{\text{input}}$, $H_{\text{o}}$, and $H_{\text{m}}$ as a function of frequency $\nu$ in Fig.~\ref{fig:Matched_Filter_Transfer_Functions}. In Fig.~\ref{fig:Matched_Filter_Transfer_Functions}c, we also plot the phase of $H_{\text{m}}$, where the phase is approximately linear with frequency for $\nu <1$ and thus preserves timing information in $s_n$. 

\begin{figure}[t!]
\begin{center}
\includegraphics[width=65mm]{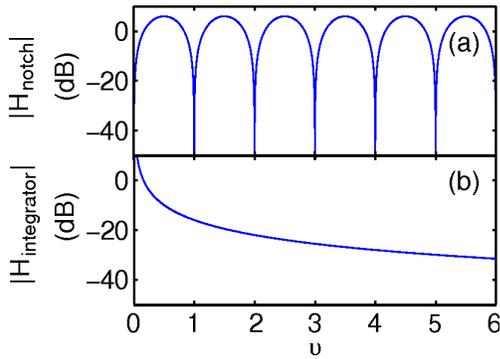}
\end{center}
\caption{\label{fig:Notch_and_Integrator_Transfer_Functions}Magnitudes of the transfer functions (a) $H_{\text{notch}}$ and (b) $H_{\text{integrator}}$.}
\end{figure}

We analyze $H_{\text{input}}$ and $H_{\text{o}}$ individually to better understand the matched filter's transfer function. We factorize $H_{\text{input}}$ into two linear operations ($H_{\text{input}} = H_{\text{notch}} H_{\text{integrator}}$), a notch filter and an integrator
\begin{equation}
H_{\text{notch}}(\nu)= \frac{\tilde{y}}{\tilde{q}}=e^{2\pi i \nu}-1,
\label{eq:H_notch}
\end{equation}
\begin{equation}
H_{\text{integrator}}(\nu)= \frac{\tilde{q}}{\tilde{v}} = \frac{1}{2 \pi i \nu},
\label{eq:H_integrator}
\end{equation}
where $\tilde{q}$ is the Fourier transform of an intermediate input-output variable and $H_{\text{notch}}$ and $H_{\text{integrator}}$ are the transfer functions of a notch filter and integrator, respectively. We plot the magnitudes of $H_{\text{notch}}$ and $H_{\text{integrator}}$ in Fig.~\ref{fig:Notch_and_Integrator_Transfer_Functions}. In Fig.~\ref{fig:Notch_and_Integrator_Transfer_Functions}a, the magnitude of the notch filter's transfer function goes to minus infinity at integer multiples of the fundamental frequency $f_{\text{o}}$. Similar types of filters have been used previously in chaotic systems to stabilize periodic orbits using continuous-time control methods.\cite{Pyragas1992, Socolar1994} The transfer function in Eq. (\ref{eq:H_integrator}) is a standard operation for integration. In Fig.~\ref{fig:Notch_and_Integrator_Transfer_Functions}b, the function $|H_{\text{integrator}}|$ diverges to infinity at $\nu = 0$ and falls off at a rate of $1/\nu$ with increasing $\nu$. When cascaded,  $H_{\text{notch}}$ and $H_{\text{integrator}}$ complement one another to form a filter that preserves low frequencies, eliminates integer multiples of $f_{\text{o}}$, and cuts out high-frequencies (see Fig.~\ref{fig:Matched_Filter_Transfer_Functions}a). 

Upon inspection of $H_{\text{o}}$ in Eq. (\ref{eq:H_dynamical}), we see that it takes on a functional form that is equivalent to the Fourier transform of the dynamical system in Eq. (\ref{eq:harmonic_oscillator_1}), given by
\begin{equation}
\frac{\tilde{u}}{\tilde{s}} = \frac{4 \pi^2 + \beta^2}{4\pi^2(1-\nu^2)+\beta(4\pi i \nu + \beta)},
\label{eq:H_system}
\end{equation}
where $\tilde{u}$ and $\tilde{s}$ are the Fourier transforms of $u(t)$ and $s(t)$, respectively. Thus, Eq. (\ref{eq:H_dynamical}) contains specific spectral information about the system's dynamics. When the transfer function $H_{\text{o}}$ is applied with $H_{\text{input}}$, it reshapes the spectrum of the transfer function for the matched filter $H_{\text{m}}$ near $\nu = 1$, as seen best in Fig.~\ref{fig:Matched_Filter_Transfer_Functions}c. Thus, the operations of the matched filter for chaos can be separated into four criteria: (i) a notch filter that eliminates the fundamental oscillation frequency $f_{\text{o}}$, (ii) an integrator that preserves the low frequencies and (iii) cuts off high frequencies falling off as $1/\nu$, and (iv) a dynamical filter that reshapes the transfer function near the fundamental oscillation frequency $f_{\text{o}}$. These four criteria are the foundation for our derivation of the pseudo-matched filter for chaos.

\section{Pseudo-matched Filter}
Our strategy for designing a pseudo-matched filter for chaos is to simplify the four criteria of the matched filter using transfer functions from components that are readily available at high-speed. In the upcoming section, we satisfy criteria (i) and (ii) using a single transfer function. We also show that criterion (iii) is more easily accomplished without an integrator. Lastly, we demonstrate that criterion (iv) is not necessary for our applications.  

\begin{figure}[h!]
\begin{center}
\includegraphics[width=65mm]{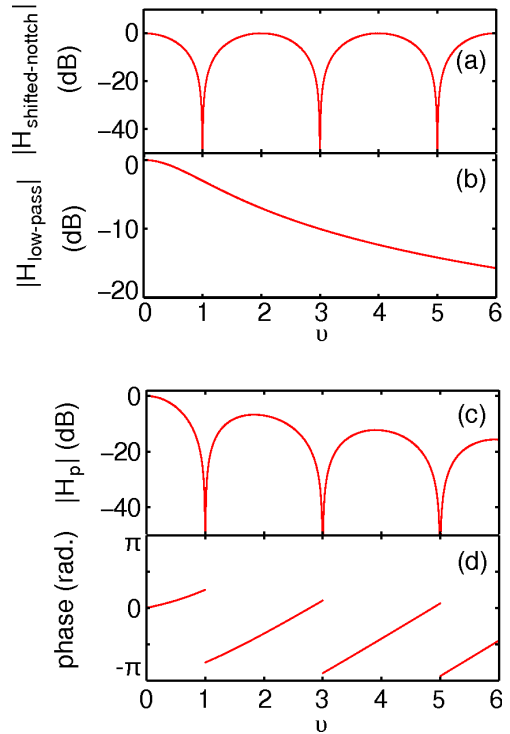}
\end{center}
\caption{\label{fig:Pseudo_Matched_Transfer_Functions}Magnitudes of the transfer functions (a) $H_{\text{shifted\text{-}notch}}$, (b)  $H_{\text{low\text{-}pass}}$, (c) $H_{\text{p}}$. In (d), the phase of $H_{\text{p}}$ is given as a function frequency.}
\end{figure}

To begin constructing our pseudo-matched filter, we select a different notch filter that is shifted in frequency but still blocks the fundamental frequency $f_{\text{o}}$. Most notch filters block integer multiples ($n = 0, 1, 2, 3, ...$) of a single frequency. Instead, we choose a notch filter that is shifted to block odd integer multiples ($2n+1 = 1, 3, 5, ...$) of a single frequency. Since the matched filter attenuates frequencies above $\nu = 1$, we conjecture that the only important spectral notch is at $f_o$, and all high-order even notches are not included in our pseudo-matched filter. The transfer function of our shifted-notch filter is

\begin{equation}
H_{\text{shifted\text{-}notch}}(\nu) = \frac{\tilde{v}_{out}}{\tilde{v}_{in}} = \frac{1}{2}(1+e^{\pi i \nu}),
\label{eq:H_shifted-notch}
\end{equation}
where $\tilde{v}_{in}$ and $\tilde{v}_{out}$ are the Fourier transforms of the input signal $v_{\text{in}}$ and output signal $v_{\text{out}}$, respectively. We plot the magnitude of $H_{\text{shifted\text{-}notch}}$ in Fig.~\ref{fig:Pseudo_Matched_Transfer_Functions}a (compare to $H_{\text{notch}}$ from Fig.~\ref{fig:Notch_and_Integrator_Transfer_Functions}a). In both plots, the fundamental frequency  $f_{\text{o}}$ is blocked. However, in Fig.~\ref{fig:Pseudo_Matched_Transfer_Functions}a, the lower frequencies ($\nu < 0.5$) are not cut. Thus, the shifted-notch filter performs two of the four operations from the matched filter; (i) it eliminates the fundamental oscillation frequency $f_o$ and (ii) preserves low frequencies. 

\begin{figure}[h!]
\begin{center}
\includegraphics[width=65mm]{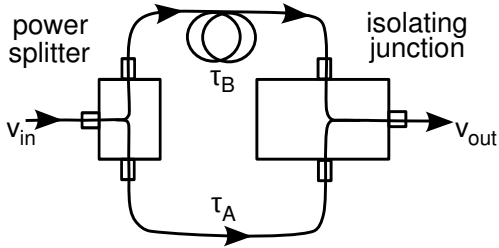}
\end{center}
\caption{\label{fig:Isolating_Junction}Pictorial realization for a high-speed ($> 1$ GHz) shifted-notch filter for voltages $v_{\text{in}}$ and $v_{\text{out}}$. An example of a broadband, high-frequency power splitter is the Mini-Circuits ZFRSC-42-S, and an example of a broadband, high-frequency hybrid-junction is the M/A-COM H-9.}
\end{figure}

In the time domain, the shifted-notch filter of Eq. (\ref{eq:H_shifted-notch}) is expressed by
\begin{equation}
v_{\text{out}}(t) = \frac{1}{2}(v_{\text{in}}(t) + v_{\text{in}}(t-\pi/\omega_{\text{o}})),
\label{eq:time_domain_shifted-notch}
\end{equation}
We compare Eq. (\ref{eq:time_domain_shifted-notch}) to Eq. (\ref{eq:matched_filter_time_domain_1}) and note that the output is no longer related to the input through a derivative. Also, the time-shift on the input signal is halved ($\pi/\omega_{\text{o}}$ instead of $2\pi/\omega_{\text{o}}$) and the shifted input $v_{\text{in}}(t - \pi/\omega_{\text{o}})$ is summed with the present state $v_{\text{in}}(t)$. In an experimental setting using high-speed electronics, where $v_{\text{in}}$ and $v_{\text{out}}$ are voltages, this shifted-notch filter can be realized using a voltage divider, time-delays (realized, for example, by coaxial cables), and an isolating hybrid junction, as illustrated in Fig.~\ref{fig:Isolating_Junction}. The lengths of the two cables used in this realization of the filter are chosen such that the difference in propagation times for electromagnetic waves to propagate through them is: $\tau_{\text{B}}-\tau_{\text{A}} = \pi/\omega_{\text{o}}$. The isolating hybrid junction sums the outputs: $v(t-\tau_{\text{A}}) + v(t-\tau_{\text{B}})$. We shift time $t \rightarrow t + \tau_{\text{A}}$ to arrive at the output signal $v_{\text{out}}$ in Eq. (14). This realization of the shifted-notch filter can scale to high-speed voltages ($> 1$ GHz).

Continuing the construction of the pseudo-matched filter, we use a first-order low-pass filter to attenuate high frequencies, rather than an integrator. We avoid the need for an integrator because the shifted-notch does not cut off low frequencies. The transfer function of the low-pass filter is 
\begin{equation}
H_{\text{low\text{-}pass}}(\nu) = \frac{\tilde{x}_{\text{out}}}{\tilde{x}_{\text{in}}} = \frac{1}{1+2 \pi i \nu/\nu_{\text{L}}},
\label{eq:low-pass transfer function}
\end{equation}
where $x_{\text{in}}$ and $x_{\text{out}}$ are the Fourier transforms of the input and output signals, respectively, and the low-pass cutoff frequency is $\nu_{\text{L}} = f_{\text{L}}/f_{\text{o}}$. We plot the magnitude of $H_{\text{low\text{-}pass}}$ in Fig.~\ref{fig:Pseudo_Matched_Transfer_Functions}b. In the figure, $H_{\text{low\text{-}pass}}$ leaves the spectral amplitude of frequencies below $\nu_L$ unchanged, while suppressing frequencies above $\nu_{\text{L}}$. Beyond $\nu = \nu_{\text{L}}$, the rate of the spectral “roll-off” of $|H_{\text{low-pass}}|$ is not $\sim \nu^{-1}$, but rather $\sim (1+\nu)^{-1}$. A first-order low-pass filter is a standard electronic component for filtering an input voltage $x_{\text{in}}$ to obtain an output voltage $x_{\text{out}}$ and satisfies approximately the third component of the matched filter criteria (iii). We note that higher-order low-pass filters (Butterworth, Chebyshev, etc.) are also available at high-speed. 

When constructing our pseudo-matched filter, we neglect the dynamical filter that reshapes the spectrum (iv). We show that using just the shifted-notch and low-pass filters allows us to achieve comparable performance to the true matched filter in a simulated radar application. Thus, we cascade the shifted-notch and low-pass filters ($H_{\text{shifted\text{-}notch}} H_{\text{low\text{-}pass}}$) to arrive at the transfer function of our pseudo-matched filter
\begin{equation}
H_{\text{p}}(\nu)= \frac{\tilde{v}_{out}}{\tilde{v}_{in}} = \frac{1+e^{\pi i \nu}}{2+4\pi i \nu/\nu_L},
\label{eq:pseudo-matched transfer function}
\end{equation}
where $\tilde{v}_{in}$ and $\tilde{v}_{out}$ are the Fourier transforms of the input and output signals of the filter, respectively. We plot the magnitude and phase of $H_{\text{p}}$ in Fig.~\ref{fig:Pseudo_Matched_Transfer_Functions}c. In the figure, the phase of $H_{\text{p}}$ is approximately linear and thus preserves timing information from $v(t)$. For comparison to the matched filter, see Fig.~\ref{fig:Matched_Filter_Transfer_Functions}c. Qualitatively, the two filters follow similar trends in both magnitude and phase. The phase in the pseudo-matched filter has a lower slope in its frequency dependence; a lower slope just constitutes a shorter time delay through the filter. However, it is clear by comparison of the magnitudes that the pseudo-matched filter is not performing the same operations as the matched filter. 

\begin{figure}[t!]
\begin{center}
\includegraphics[width=80mm]{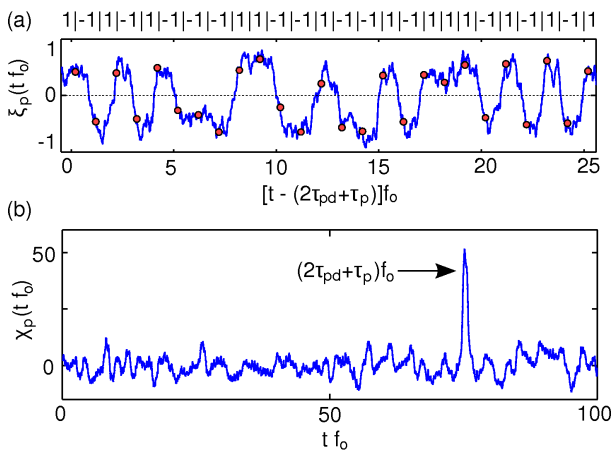}
\end{center}
\caption{\label{fig:Pseudo_Matched_Output}Output of pseudo-matched filter. (a) Time series of the output of the matched filter $\xi_{\text{p}}(t)$ (blue) while driven by $v(t) = u(t) + AWGN$ ($SNR = -5.9$ dB). The signal $\xi_{\text{p}}(t)$ is sampled with uniform spacing (red dots) at a clock frequency $f_{\text{o}} = \omega_{\text{o}}/2\pi$. Above the waveform, a single-bit discrete sampling of the waveform is shown. From the figure, we see that all of the relevant information from $s_n$ is encoded in $\xi_{\text{p}}(t)$. (b) The switching state $s(t)$ is sampled and $s_n$ is stored for $N=100$. The output of the pseudo-matched filter $\xi_{\text{p}}(t)$ drives Eq. (\ref{eq:correlation_pseudo}) and the output $\chi_{\text{p}}(t)$ peaks at time $t = 2\tau_{\text{pd}}+\tau_{\text{p}}$.}
\end{figure}

We now apply the pseudo-matched filter to the chaotic waveform generated by Eqs. (\ref{eq:harmonic_oscillator_1})-(\ref{eq:harmonic_oscillator_2}) and examine its output in the time-domain. We drive the pseudo-matched filter with $v(t) = u(t) + AWGN$, where $v(t)$ has a $SNR$ of $-5.9$ dB (see Fig.~\ref{fig:Matched_Filter_Time_Series}c). In Fig.~\ref{fig:Pseudo_Matched_Output}a, we plot the pseudo-matched filter's output $\xi_{\text{p}}(t)$. From the figure, we see that pseudo-matched filter has effectively removed the main oscillation frequency $f_{\text{o}}$ and what remains is a digital-like signal $\xi_{\text{p}}(t)$, where the noise has also been reduced. We note that a considerable amount of noise is still present in comparison to Fig.~\ref{fig:Matched_Filter_Time_Series}d. Using a correlation between the original $s(t)$ and $\xi_{\text{p}}(t)$, we determine the time delay through the pseudo-matched filter $\tau_{\text{p}}$ to be approximately $0.14/f_{\text{o}}$. We compensate for the delay, and sample $\xi_{\text{p}}(t)$ at $f_{\text{o}}$, assigning binary values using the relation: $-1$ if $\xi_{\text{p}} (t_n) \leq 0$ and $+1$ if $\xi_{\text{p}} (t_n) > 0$, where $t_n$ is the $n^{\text{th}}$  sampling time. In Fig.~\ref{fig:Pseudo_Matched_Output}a, we see that, with this particular $SNR$, the discrete sampling of $\xi_{\text{p}}(t)$ is equivalent to $s_n$ from Fig.~\ref{fig:Matched_Filter_Time_Series}b. 

Next, we parallel the construction of the time delay tap from Eq. (\ref{eq:correlation_matched}) for the output of the pseudo-matched filter. In this case, the time delay tap is
\begin{equation}
\chi_{\text{p}}(t) = \sum\limits_{n=1}^N s_n \xi_{\text{p}}(t-\tau_n),
\label{eq:correlation_pseudo}
\end{equation}
where $\chi_{\text{p}}(t)$ is the output of the time-domain correlation. We plot an example of $\chi_{\text{p}}(t)$ in Fig.~\ref{fig:Pseudo_Matched_Output}b using the the same chaos and $s_n$ that were used to calculate $\chi_{\text{m}}(t)$ in Fig.~\ref{fig:Tapped_Delay_Line}b. By visually comparing $\chi_{\text{p}}(t)$ to $\chi_{\text{m}}(t)$, we see that the output correlation peaks are qualitatively similar, but $\chi_{\text{p}}(t)$ has more noise. In the remaining section, we establish criteria for quantitatively comparing these correlation waveforms and use these criteria to weigh each filter's performance.  

\section{Matched vs. Pseudo-matched}

In radar applications, the ability to correctly identify the location of the correlation peak in the correlation operation is the useful measure. Therefore, in order to compare quantitatively the output from the matched filter and pseudo-matched filter, we weigh each filter's performance based on the peak width and output $SNR$ of $\chi_{\text{m,p}}(t)$. We also present an approximate analytical form for each correlation's output $SNR$.

The peak widths of the output-correlation functions give the resolutions of each radar system. We measure $\Delta_{\text{m}}$ and $\Delta_{\text{p}}$, the full-width at half maximum (FWHM) time of the correlation output peaks using the matched and pseudo-matched filters, respectively. For the most ideal measure of each filter's correlation peak width, we measure $\Delta_{\text{m,p}}$ in cases where no noise is present in the received waveform $v(t) = u(t)$. We note that these widths are independent of $N$, the number of stored data points in the correlation calculation of Eqs. (\ref{eq:correlation_matched}) and (\ref{eq:correlation_pseudo}). Using a Gaussian fit to the peak of $\chi_{\text{m,p}}(t)$, we obtain peak widths $\Delta_{\text{m}} f_{\text{o}} = 0.55$ and $\Delta_{\text{p}} f_{\text{o}} = 0.73$ (see Appendix). Using these values of $\Delta_{\text{m,p}}$ and scaling $f_{\text{o}}$ to 1 GHz, we calculate the theoretical resolutions of the matched and pseudo-matched filters to be 0.17 m and 0.22 m, respectively. In this example, the ranging resolutions differ by 5 cm. Thus, this is not a critical difference for radar applications that localize targets like planes or cars, and the pseudo-matched filter has an acceptable ranging resolution in comparison to the matched filter. 

\begin{figure}[t!]
\begin{center}
\includegraphics[width=80mm]{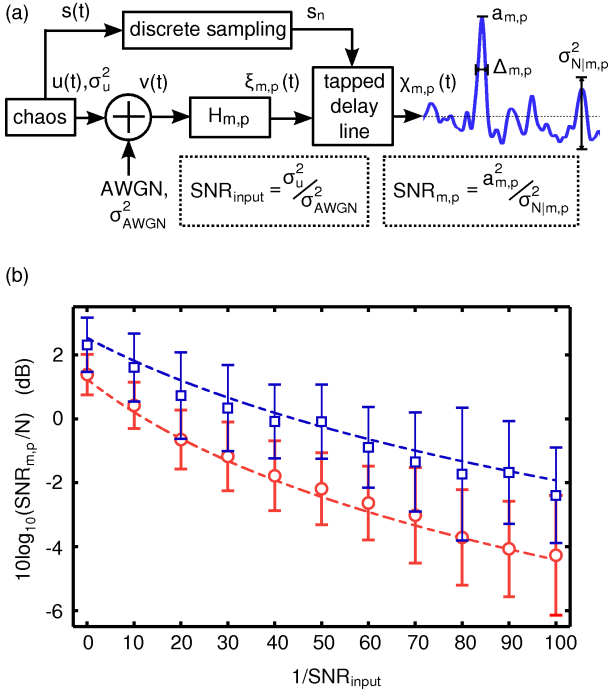}
\end{center}
\caption{\label{fig:SNR_output}(a) Block diagram for testing the matched and pseudo-matched filters in a simulated radar application. (b) Output-correlation $SNR$'s of the matched (blue $\square$) and pseudo-matched (red $\bigcirc$) filters scaled by $N$ on a logarithmic scale as a function of $1/SNR_{\text{input}}$. For each value of $SNR_{\text{input}}$, 100 calculations of $SNR_{\text{m,p}}$ were performed using sequence of $s_{n}$ for $n = n_{\text{o}}$ to $n = n_{\text{o}} + N$, where $n_{\text{o}}$ is a random positive integer and $N = 50$. The mean value of the calculated $SNR_{\text{m,p}}$ is plotted with the respective standard deviations. The bue and red dotted lines give the theoretical predictions of the $SNR_{\text{m,p}}$ as references for the matched and pseudo-matched filters, respectively. Cases for larger $N$ were verified to have quantitatively similar results.}
\end{figure}

Next, we measure the output $SNR$'s of the matched and pseudo-matched filter correlations using the correlation peak heights $a_{\text{m,p}}$ and the surrounding correlation noise floors. The output $SNR$ in $\chi_{\text{m,p}}(t)$ is 
\begin{equation}
SNR_{\text{m,p}} = \frac{a_{\text{m,p}}^2}{\sigma_{\text{N}|\text{m,p}}^2},
\label{eq:SNR_Measure}
\end{equation}
where $a_{\text{m,p}}$ is the peak height of $\xi_{\text{m,p}}(t)$ from a Gaussian fit (see Appendix) and $\sigma_{\text{N}|\text{m,p}}^2$ is the output variance of the correlation noise floor (note that the mean of the noise floor $\sim 0$) for the matched and pseudo-matched filters, respectively. We present a summary of these quantities in the block diagram shown in Fig.~\ref{fig:SNR_output}a. In the diagram, we also review the waveforms and processes used for generating $\xi_{\text{m,p}}(t)$ and $\chi_{\text{m,p}}(t)$ and highlight the two relevant quantities, $SNR_{\text{input}}$ and $SNR_{\text{m,p}}$. We calculate $SNR_{\text{m,p}}$ as a function of the input $SNR_{\text{input}}$. The results of these calculations are given in Fig.~\ref{fig:SNR_output}b. In addition, we use the distributions from $\xi_{\text{m}}(t)$ and $\xi_{\text{p}}(t)$ from the two different cases $v(t) = AWGN$ and $v(t) = u(t)$ to derive a an analytical prediction for $SNR_{\text{m,p}}$ (see Appendix for derivations). These theoretical predictions are plotted with dotted lines in Fig.~\ref{fig:SNR_output}b. These plots represent the performances of the matched and pseudo-matched filters in a simulated radar. 

From Fig.~\ref{fig:SNR_output}b, it is clear that the matched filter outperforms the pseudo-matched filter in the output $SNR$ of a radar correlation. Without noise in the system, the matched and pseudo-matched filters perform with output correlation $SNR$'s of $2.6+10\text{log}_{10}(N)$ dB and $1.3+10\text{log}_{10}(N)$ dB, respectively. For $SNR_{\text{input}}$ = 1/100, the output $SNR$'s decrease to $-2.0+10\text{log}_{10}(N)$ dB and $-4.4+10\text{log}_{10}(N)$ dB, respectively. In Fig.~\ref{fig:SNR_output}b, the average difference between $SNR_{\text{m}}$ and $SNR_{\text{p}}$ is 2.0 dB. We note that this difference is independent of $N$ and therefore fully characterizes the filter performances. Thus, where small loss is acceptable in the performance of the radar, the pseudo-matched filter is a simpler alternative to the system's analytically matched filter for chaos. 

As a final example, we use the theoretical $SNR_{\text{m,p}}$ to predict when the matched and pseudo-matched filters will fail in a radar application. Failure occurs when $SNR_{\text{m,p}}$ falls below a certain threshold. For a radar system that is capable of storing $N$ = 50 data points and has a desired output correlation $SNR$ of 33 dB, the matched and pseudo-matched filters will fail at a $1/SNR_{\text{input}}$ of approximately $25$ and $70$, respectively. If, in this application, the input $SNR$ is such that $1/SNR_{\text{intput}} < 10$, then a radar with either the matched or pseudo-matched filter will be able to range, on average, without failure. The choice between the matched and pseudo-matched filter is therefore an application-dependent problem, and, as the bandwidth of this system scales higher, one must also begin to weigh each filter's high-frequency capabilities as well as its baseline performance. 

\section{CONCLUSIONS}
In conclusion, for the chaotic system presented in Ref. [\onlinecite{Corron2010}], we derive empirically a sub-optimal filter for removing noise from the waveforms generated by this dynamical system. This sub-optimal filter performs approximately three out of four of the linear operations from the matched filter: (i) eliminates the fundamental oscillation frequency $f_{\text{o}}$, (ii) preserves the low frequencies, and (iii) cuts off high frequencies. Our filter, deemed a pseudo-matched filter, is composed of a shifted-notch filter and a first-order low-pass filter. In the context of a radar concept that uses a time delay tap as a correlation measure, we have shown that, depending on the application, the pseudo-matched filter may be an acceptable substitute for the matched filter. In addition, we acknowledge that the psuedo-matched filter can be further improved using higher order low-pass filters and additional shifted-notch filters. We present this current version of the pseudo-matched filter to illustrate our method for its derivation and emphasize its simplicity.

Lastly, we note that our analysis highlights the flexibility of Corron \textit{et al.}'s findings. The chaos from the dynamical system in Eqs. (\ref{eq:harmonic_oscillator_1}) - (\ref{eq:harmonic_oscillator_2}) can be processed by a linear filter to recover an underlying digital waveform. We capitalize on this system's elegance to create a pseudo-matched filter. Although it is less optimal when compared to the matched filter, the pseudo-matched filter shows that the advantages of this chaotic system can be adapted for applied settings that use commercially available, high-speed filters.

\section{ACKNOWLEDGEMENTS}
We gratefully acknowledge Damien Rontani for useful discussions, G. Martin Hall with help in radar concepts, and the financial support of Propagation Research Associates (PRA) Grant No. W31P4Q-11-C-0279.

\appendix
\section{Additive Noise}

Because the output from MATLAB's ODE45 uses a variable timestep, we resample $u(t)$ using a linear interpolation with time steps $\delta_t$, where $\delta_t f_{\text{o}} = 10^{-2}$. To simulate environmental noise, we add noise to the waveform $u(t)$ using random numbers spaced by time units $\delta_t$. The random numbers are calculated from a Gaussian distribution with zero mean. For different points along the $1/SNR_{\text{input}}$ axis of Fig.~\ref{fig:SNR_output}b, the variance of the $AWGN$ is varied accordingly.

\section{Gaussian Fits}
We measure the correlation peak width and height using a Gaussian fit
\begin{equation}
f(t) = a_{\text{m,p}}e^{(t-(2\tau_{\text{pd}}+\tau_{\text{m,p}}))^2/2c_{\text{m,p}}^2},
\label{eq:gaussian_fit}
\end{equation}
where $a_{\text{m,p}}$ and $c_{\text{m,p}}$ are free parameters that are fit to the correlation peak heights and widths. Using $f(t)$ to fit $\chi_{\text{m,p}}(t)$, we obtain a FWHM peak width $\Delta_{\text{m,p}} f_{\text{o}} = 2\sqrt{2ln(2)}c_{\text{m,p}}$ and peak height $a_{\text{m,p}}$.

\section{Analytical SNR's}

We derive analytical forms for the output-correlation $SNR$ of the matched and pseudo-matched filters. To do so, we approximate Eq. (\ref{eq:SNR_Measure}) as
\begin{equation}
SNR_{\text{m,p}} = \frac{a_{\text{m,p}}^2}{\sigma_{\text{N}|\text{m,p}}^2} \sim \frac{(A_{\text{m,p}} N)^2}{\sigma_{1|\text{m,p}}^2 N +\sigma_{2|\text{m,p}}^2 N},
\label{eq:SNR_Theory}
\end{equation}
where $A_{\text{m,p}}$ is a constant that characterizes the growth rate of the correlation peak height with $N$, $\sigma_{1|\text{m,p}}^2$ is a constant determined in the noise-free case where $v(t) = u(t)$, and $\sigma_{2|\text{m,p}}^2$ is a function of $SNR_{\text{input}}$ in the case where $v(t) = AWGN$. Recall that the numerators and denominators of Eq. (\ref{eq:SNR_Theory}) represent the power of peak heights of the correlation and the surrounding noise floor, respectively. We derive each of the three terms $A_{\text{m,p}}$, $\sigma_{1|\text{m,p}}^2$, and $\sigma_{2|\text{m,p}}^2$ in the following sections.

The correlation peak heights for the matched and pseudo-matched filters grow at different rates. In the correlation operations of Eqs. (\ref{eq:correlation_matched}) and (\ref{eq:correlation_pseudo}), the peaks occur at times $t_{\text{m,p}}^* = 2\tau_{\text{pd}}+\tau_{\text{m,p}}$ for the matched and pseudo-matched filters, respectively. At time $t^*_{\text{m,p}}$, $\xi_{\text{m,p}}(t)$ is aligned with $s_n$ and the output correlation is
\begin{equation}
\chi_{\text{m,p}}(t^*_{\text{m,p}}) \sim \sum\limits_{n=1}^N |\xi_{\text{m,p}}(t^*_{\text{m,p}}-\tau_n)| \sim A_{\text{m,p}}N.
\label{eq:Peak_Times}
\end{equation}
We approximate $A_{\text{m,p}}$ from the local maxima of $|\xi_{\text{m,p}}(t)|$. To do so, we examine the noise-free case where $v(t) = u(t)$ and collect a subset of points $|\xi_{\text{m,p}}(t_{\text{m,p}}^{(r)})|$ where $t_{\text{m,p}}^{(r)}$ are the times of local maxima in $|\xi_{\text{m,p}}(t)|$. We average $|\xi_{\text{m,p}}(t_{\text{m,p}}^{(r)})|$ to obtain $A_{\text{m}} = 0.67$ and $A_{\text{p}} = 0.51$ using a time-length $t f_{\text{o}} \sim 10^4$.

To approximate the value of $\sigma_{1|\text{m,p}}^2$, we also examine $\xi_{\text{m,p}}(t)$ in the noise-free case where $v(t) = u(t)$. The deterministic noise floor in a correlation measurement is also known as its side-lobes; the side-lobes result from non-zero contributing terms in the correlation $\chi_{\text{m,p}}(t)$ when $t \neq t_{\text{m,p}}^*$. Using the central limit theorem, we approximate the variance of these nonzero terms as $\sigma_{1|\text{m,p}}^2N$, where $\sigma_{1|\text{m,p}}^2$ is the variance of the signal $\xi_{\text{m,p}}(t)$. In this approximation, we find that $\sigma_{1|\text{m}}^2 = 0.25$ and $\sigma_{1|\text{p}}^2 = 0.20$. 

It remains to calculate the contributions to the noise floor of the correlation from additive noise. To do so, we examine $\xi_{\text{m,p}}(t)$ in the case where $v(t) = AWGN$. Similar to the case for the side-lobes, we use the central limit theorem to approximate the contribution of the $AWGN$ to the noise floor of the correlation as $\sigma_{2|\text{m,p}}^2N$, where $\sigma_{2|\text{m,p}}^2$ is the variance of the signal $\xi_{\text{m,p}}(t)$. However, $\sigma_{2|\text{m,p}}^2$ depends on the variance of the input $AWGN$
\begin{equation}
\sigma_{2|\text{m,p}}^2 = \alpha_{\text{m,p}}\sigma_{AWGN}^2,
\label{eq:Noise_Relation}
\end{equation}
where $\alpha_{\text{m,p}}$ is the noise attenuation factor of the matched and pseudo-matched filters, respectively. We measure the values $\alpha_{\text{m}} = 1/289$ and $\alpha_{\text{p}} = 1/255$. Lastly, we use that $\sigma_{AWGN}^2 = \sigma_{u}^2/(SNR_{\text{input}})$ to rewrite Eq. (\ref{eq:SNR_Theory}) as
\begin{equation}
SNR_{\text{m,p}} \sim N\frac{A_{\text{m,p}}^2}{\sigma_{1|\text{m,p}}^2 +\alpha_{\text{m,p}}\frac{\sigma_{u}^2}{SNR_{\text{input}}}},
\label{eq:SNR_Theory_2}
\end{equation}
where $\sigma_{u}^2 = 1.34$ is the power of the chaotic signal $u(t)$. We plot Eq. (\ref{eq:SNR_Theory_2}) as a function of $1/SNR_{\text{input}}$ for the matched and pseudo-matched filters in Fig.~\ref{fig:SNR_output}b.

\bibliographystyle{apsrev4-1.bst}
\small 
\bibliography{references_Pseudo}

\end{document}